\documentclass[aps,prb,twocolumn,groupedaddress,showpacs]{revtex4}

\usepackage{graphicx}

\begin{document}

\title{Raman signature of electron-electron correlation in chemically doped few-layer graphene}

\author{Matteo Bruna}
\author{Stefano Borini}
\email[]{s.borini@inrim.it} \affiliation{INRIM, Strada delle Cacce
91, I-10135 Torino (Italy)}

\begin{abstract}

We report an experimental Raman study of few-layer graphene after
chemical doping achieved by a plasma process in CHF$_3$ gas. A
systematic reduction of both the splitting and the area of the $2D$
band is observed with increasing the doping level. Both effects can
be ascribed to the electron-electron correlation, which on the one
hand reduces the electron-phonon coupling strength, and on the other
hand affects the probability of the double resonant Raman process.

\end{abstract}

\maketitle

Since the possibility to isolate and identify graphene atomic layers
has been experimentally demonstrated,~\cite{Novoselov2005} the
direct observation of many peculiar physical phenomena (e.g. the
half-integer quantum hall effect for Dirac fermions
~\cite{NovoselovKim2005}) has become accessible by standard
characterization techniques, such as charge transport measurements,
vibrational spectroscopy, scanning probe spectroscopy. In particular
Raman spectroscopy, which is a very powerful technique for studying
graphene, able to easily discriminate monolayers from bilayers and
trilayers,~\cite{Ferrari2006} has unveiled many important features
of the graphene system, such as Kohn anomalies in the phononic
spectrum,~\cite{Yan2007, Das2008, Mafra2009} and the failure of the
adiabatic Born-Oppenheimer approximation in describing the
electron-phonon coupling (EPC) for Brillouin-zone center optical
phonons.~\cite{Pisana2007, Lazzeri2006} Here we show that even the
evolution of electron-electron correlation can be experimentally
revealed in the Raman spectrum of multilayer graphene, with a
doping-dependent hallmark in the $2D$ band.

The $2D$ band arises from a double resonant Raman process, where the
intervalley scattering of two electrons is accompanied by the
emission of two phonons with opposite momentum (around the K point
of the Brillouin zone).~\cite{Ferrari2006} Therefore, such a Raman
process is very sensitive to the electronic bands, which determine
all the possible initial and final states for the electrons involved
in the intervalley scattering, allowing to monitor the evolution of
the electronic band structure with the number of stacked graphene
layers (from monolayer to bilayer to few-layer). Very recently, it
was shown that even the change in the band structure due to a
different stacking order can be captured in the Raman $2D$ band of
few-layer graphene.~\cite{Lui2011}

Moreover, the Raman spectrum of graphene is affected by EPC, which,
for optical phonons near the $K$ point, shows up in the dispersive
behavior of the $D$ and $2D$ band. Indeed, the dependence of the
Raman $2D$ band on the excitation energy is proportional to the
slope of the phononic band near $K$, which is determined by the EPC
strength.~\cite{Piscanec2004} In such a scenario, the
electron-electron interactions play an important role, giving a
major contribution to the dispersion of the highest optical phonon
branch near $K$ in neutral graphene.~\cite{Lazzeri2008} In fact, the
experimental phonon slope can be theoretically reproduced by ab
initio calculations only within the $GW$ approach, where nonlocal
exchange-correlation effects are included. Recently, Attaccalite et
al.~\cite{Attaccalite2010} theoretically showed that the deformation
potential (i.e. the EPC strength) for optical phonos near $K$ is
strongly affected by the charge carriers density in graphene, due to
electron-electron correlation effects, so that the $D$ and $2D$
Raman bands should reflect such a doping dependence.

We carried out an experimental study of chemically-doped few-layer
graphene, monitoring the Raman $2D$ band as a function of the doping
level. Graphene was mechanically exfoliated from natural graphite
and deposited on 285 nm thick SiO$_2$ on Si substrates. The number
of stacked graphene layers in the deposited flakes was established
by an optical contrast analysis,~\cite{Bruna2009} using an optical
microscope and appropriate bandpass filters, so that up to 7 stacked
layers were clearly distinguished.~\cite{Bruna2010} Then, chemical
doping was performed by a radio frequency (RF) plasma process in
CHF$_3$ gas, at RF power of 15 W and gas pressure of 100 mTorr. We
have shown elsewhere~\cite{Bruna2010} that, at such experimental
conditions, a high p-type doping of graphene can be achieved, due to
the adsorption of fluorine atoms at the graphene surface. Moreover,
no structural modification of the graphene lattice occurs, as
indicated by the absence of the Raman $D$ peak. Raman spectra were
acquired by means of a Jobin-Yvon U1000 Raman spectrometer equipped
with a microscope (100X objective) and with an Ar-Kr laser, using
the excitation wavelength $\lambda=514.5$ nm. The incident laser
power focused on the sample was adjusted to be less than 5 mW, in
order to avoid any local heating effect.

\begin{figure*}
  \includegraphics{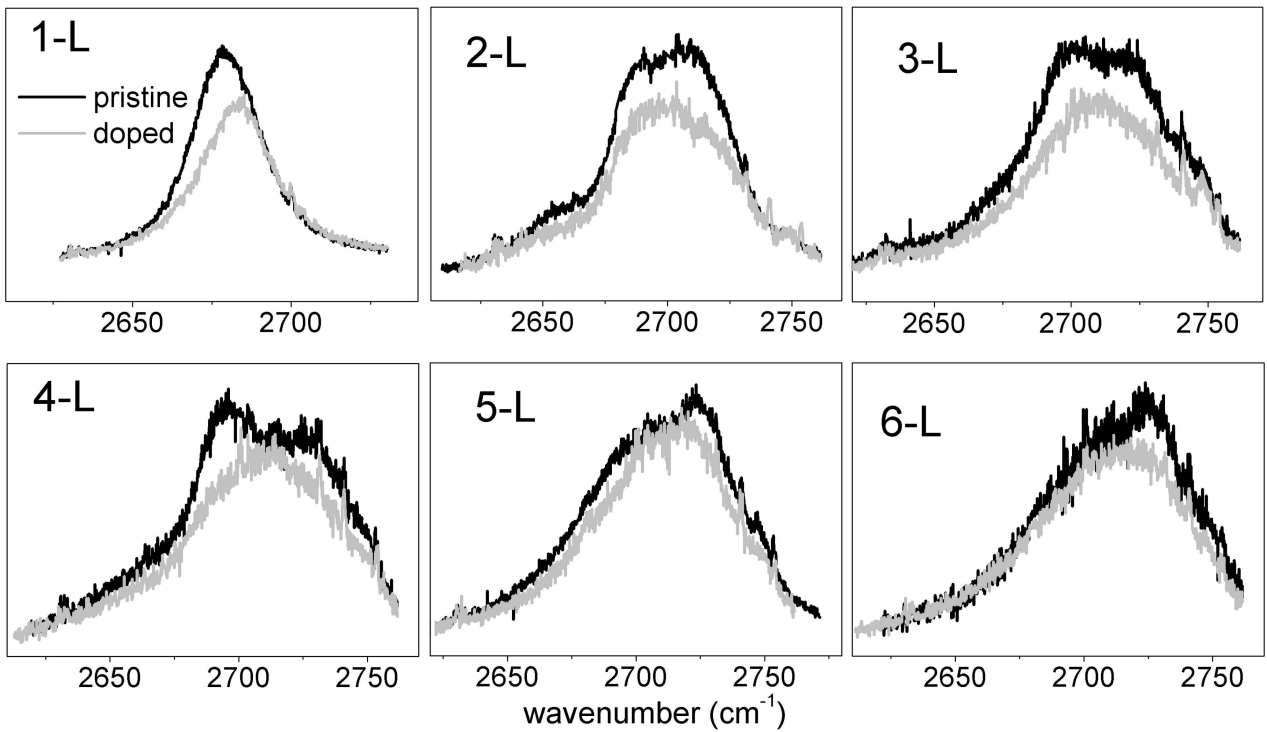}%
  \caption{Raman 2D band measured on various graphene flakes on the same substrate,
before black line) and after (grey line) the chemical doping
process. The label N-L, with N=1-6, indicates the number of stacked
graphene layers in the analyzed flake.}\label{f1}
\end{figure*}

In Fig.~\ref{f1} the change induced by chemical doping in the 2D
band of N-layer graphene (N=1-6) is reported. The spectra were taken
on the same substrate, before and after the plasma treatment. The
result was checked on different flakes on the same substrate and on
different substrates.
On the one hand, as expected for high p-type doping, the monolayer
spectrum is blue-shifted, consistently with the
literature.~\cite{Das2008} On the other hand, all the few-layer
spectra display a common feature, i.e. a systematic reduction of the
band splitting, which was theoretically predicted by Attaccalite et
al.~\cite{Attaccalite2010} by taking into account electron-electron
correlation effects in the calculation of the deformation potential
as a function of doping. Indeed, the $2D$ band splitting in
multilayer graphene is directly related to the slope of the highest
optical phonon branch near K, i.e. to the EPC strength, which is
tuned by the charge carrier density. In our experiments we always
observed that the sub-peaks composing the few-layer $2D$ band tend
to shrink toward a single spectral position at high doping levels,
indicating a strong decrease of the phonon energy dispersion.
Moreover, even the $2D$ band area seems to be systematically
decreased by the chemical doping, as already reported for monolayer
graphene.~\cite{Das2008, Basko2009}

\begin{figure}
  \includegraphics{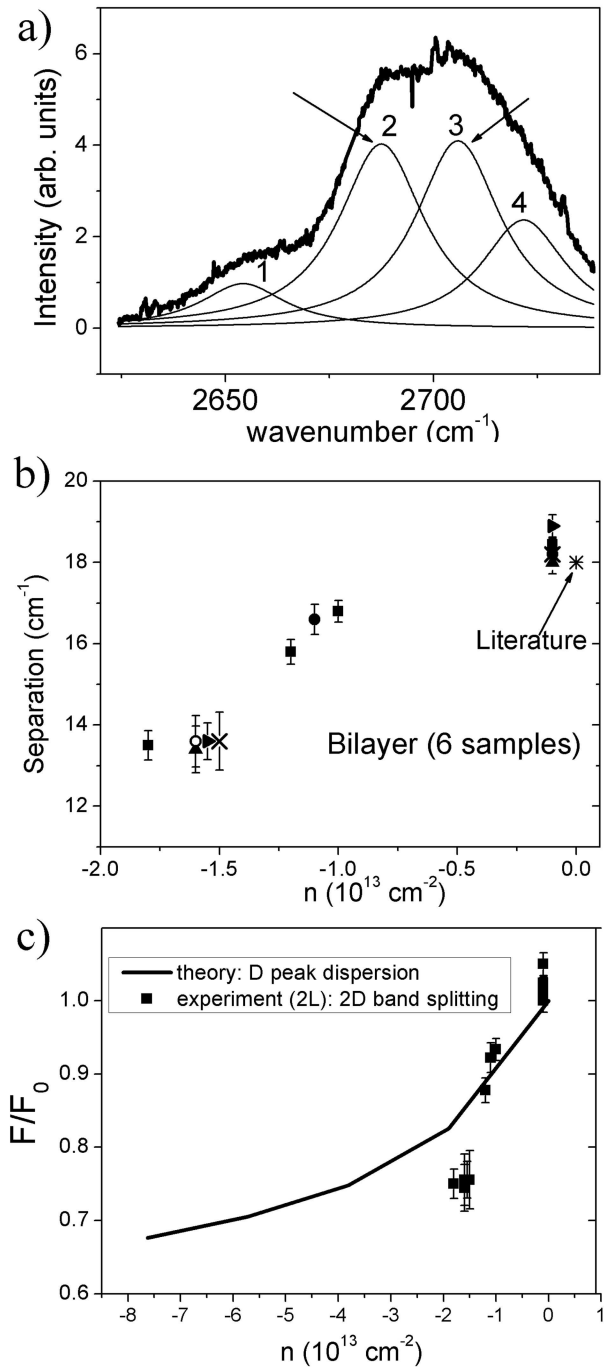}%
  \caption{Analysis of the spectral separation of the bilayer $2D$ sub-peaks,
labeled as "2" and "3" in (a), as a function of the doping level as
estimated in Ref.~\cite{Bruna2010} (b). The asterisk shows for
comparison a typical value reported in the literature
~\cite{Malard2007} for undoped bilayer graphene. In (c), the
relative variation of the splitting shown in (b) is compared with
the relative variation of the $D$ peak dispersion slope
theoretically predicted by Attaccalite et
al.~\cite{Attaccalite2010}}\label{f2}
\end{figure}

In order to confirm these qualitative observations, we have analyzed
in detail the evolution of the bilayer $2D$ band with the doping
level. In fact, in the bilayer case the $2D$ band is described by
the convolution of four lorentzian peaks, corresponding to the four
possible resonant processes giving rise to the Raman signal, and
among the four peaks, two of them (labeled as 2 and 3 in
Fig.~\ref{f2}a) are known to be prominent.~\cite{Ferrari2006}

Therefore, we have fitted the bilayer spectra by four lorentzian
peaks with FWHM fixed at 25 cm$^{-1}$ (typical width of the single
$2D$ peak of monolayer graphene),~\cite{Malard2007} and we have
analyzed the evolution of the two most intense peaks with the doping
level. The peaks width was assumed to be unaffected by the doping,
on the basis of the observation of the monolayer 2D peak width, which did
not display any significant variation after the plasma treatment.

As shown in Ref.~\cite{Bruna2010}, the modification induced by the
plasma treatment was not stable under ambient conditions, and the
Raman spectra changed with the passing of time, slowly tending to
their pristine form. Therefore, we were able to gradually vary the
doping level on top of the graphene flakes, monitoring the Raman
spectrum as a function of doping. Moreover, we could estimate the
carrier density by means of the analysis of the $G$ band splitting
carried out in Ref.~\cite{Bruna2010}, in order to plot the peaks
position as a function of the doping level. Indeed, the bilayer $G$
band is splitted in two peaks by heavy top-layer doping, and the
behavior as function of the carrier density can be fitted by the
thoretical curves given in Ref. ~\cite{Gava2009}, yielding to an
estimate of the doping level in the experiment.

In Fig.~\ref{f2}b the reduction of the spectral separation of peaks
2 and 3 with increasing the doping level can be clearly appreciated.
The reported data were obtained by fitting the $2D$ band of six
bilayer samples, and error bars were estimated by propagating the
peaks position standard deviations given by the fitting procedure. A
typical peaks separation value reported in the literature
~\cite{Malard2007} is also shown as a reference in Fig.~\ref{f2}b.
The robustness of the fit results was checked by repeating the
fitting procedure with the FWHM as a free parameter (i.e. not fixed
at the value of 25 cm$^{-1}$), in order to take into account
possible variations of the peaks width with the doping. Although a
higher uncertainty for the peaks separation values was obtained in
this case, a splitting reduction of about the same magnitude could
still be clearly appreciated, confirming the results reliability.

The consistence of this observation with theory
~\cite{Attaccalite2010} was checked by comparing the experimental
peaks separation (normalized by the value reported in the literature
for the undoped case) to the theoretically predicted $D$ peak
dispersion slope (normalized by the its value at zero doping).
Indeed, since the $2D$ band splitting is proportional to the $D$
peak dispersion, the relative variation of the two quantities is
expected to be the same. In Fig.~\ref{f2}c the direct comparison of the
two curves is reported (where F is either the experimental peaks separation or the theoretical $D$ peak dispersion slope, and F$_{0}$ the corresponding value at zero doping), suggesting a qualitative agreement between
theory and experiment. More experiments carried out within a broader
range of carrier density would be necessary to definitely quantify
the doping effect on the $2D$ band splitting. However, the chemical
doping approach here reported gives rise to a high carrier density
($> 10^{13}$ cm$^{-2}$) which cannot be reached in standard field
effect experiments. In order to further increase the doping level,
an electrochemical gating approach should be
employed,~\cite{Das2008} whose experimental realization is more
challenging.

\begin{figure}
  \includegraphics{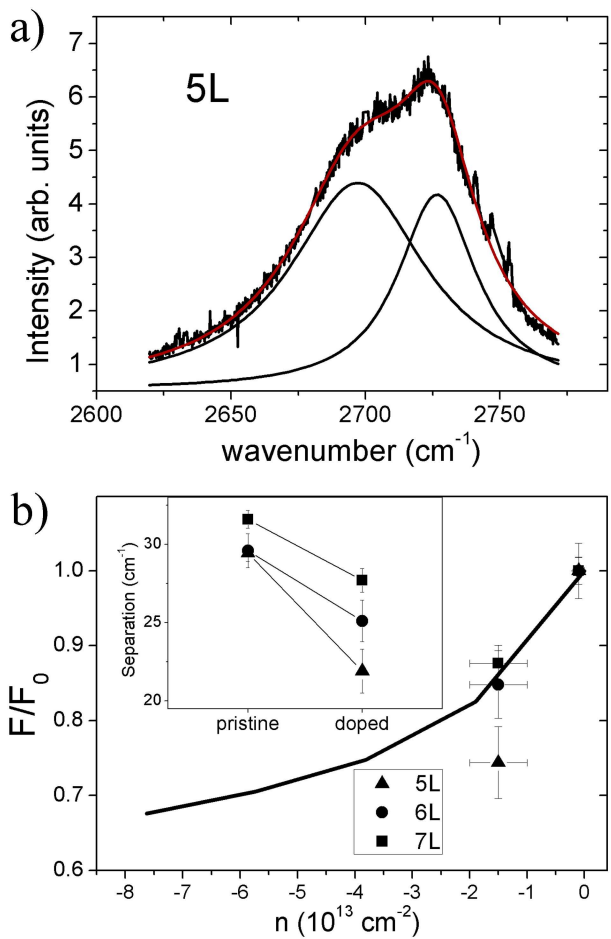}%
  \caption{(color online). (a) Fitting of the 5-layer 2D band by two lorentzian peaks;
(b) spectral separation of the two peaks in pristine and doped
samples (inset), and comparison of the experimental spectral
separation (symbols) to the theoretical $D$ peak dispersion slope
(line) after normalization by their value at zero doping.}\label{f3}
\end{figure}

Moreover, an analogous study was carried out on 5-layer, 6-layer and
7-layer graphene samples, in order to corroborate the results of the
bilayer analysis. Indeed, with increasing the number of stacked
layers, the few-layer graphene Raman spectrum approaches the bulk
graphite spectrum, where the 2D band is well fitted by two
lorentzian peaks. In Fig.~\ref{f3} the results of fitting the
experimental spectra by two lorentzian peaks with variable FWHM are
reported, and the evolution of the peaks separation with increasing
the doping level is plotted, showing a clear reduction as in the
bilayer case. The magnitude of the effect is comparable to the
bilayer case, and of the same order of the theoretical prediction.
Horizontal error bars are due to the uncertainty of the doping
value, because in this case it was not possible to exploit the $G$
band analysis to extract the carrier density as in the bilayer case.
However, basing on the values obtained for monolayers and bilayers,
it is likely that the doping value immediately after plasma
treatment ranges between $1$ and $2\times10^{13}$ cm$^{-2}$ for
every few-layer graphene.

Furthermore, the area of the $2D$ band was analyzed as well, in
order to study its dependence on chemical doping. For monolayer
graphene, it has been shown that the $2D$ band intensity is
decreased by doping,~\cite{Das2008, Basko2009} due to
electron-electron scattering processes which affect the resonant
Raman process. Thus, we have performed the same analysis of
Ref.~\cite{Basko2009} on the $2D$ band of few-layer graphene, using
the area of a monolayer $G$ band (measured on the same substrate and
in the same experimental conditions) as a normalization factor, in
order to neglect spurious experimental contributions to the measured
intensity variation. Indeed, the $G$ band area is almost unaffected
by doping in monolayer graphene as far as the Fermi level $E{_F}<<1$
eV,~\cite{Basko2009} whereas in the bilayer case it can display
strong modifications due to inversion symmetry
breaking.~\cite{Bruna2010, Malard2008, Yan2009, Zhao2010, Ando2009}

\begin{figure}
  \includegraphics{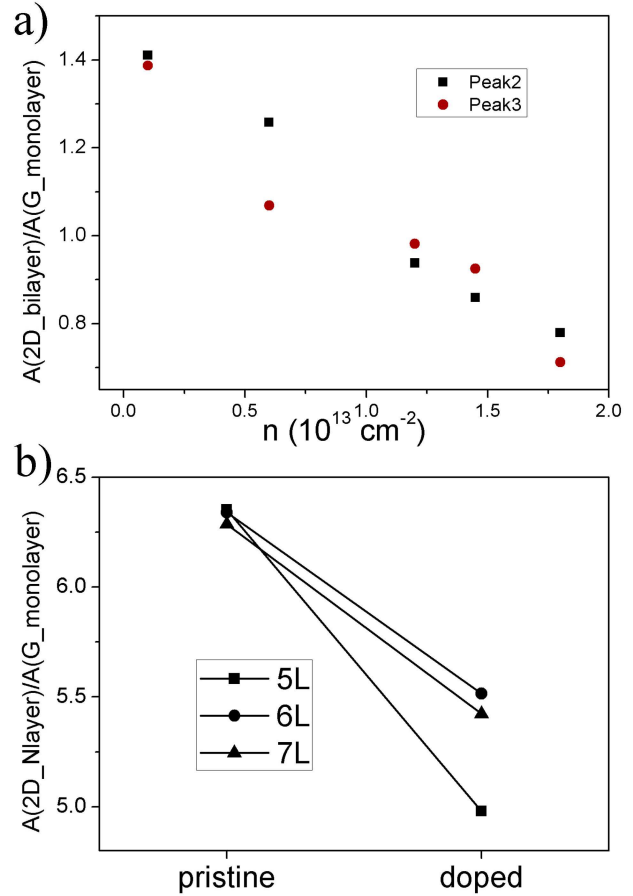}%
  \caption{(color online). Behavior of the normalized area of the bilayer
$2D$ band sub-peaks (a) and of the 5-L, 6-L and 7-L $2D$ band (b) as
a function of the doping level.}\label{f4}
\end{figure}

Fig.~\ref{f4} shows an evident systematic reduction of all the
considered areas, with a decrease of about $50\%$ for the bilayer
peaks with an estimated doping level of less than $2\times10^{13}$
cm$^{-2}$.

Such a result strongly suggests that the probability of the Raman
process involving optical phonons near $K$ in few-layer graphene is
reduced by the onset of doping-induced electron-electron scattering
processes, in analogy with the monolayer case.

In summary, we have experimentally observed the reduction of both
the splitting and the area of the Raman $2D$ band with increasing
the doping level in N-layer graphene, for N ranging from 2 to 7. The
analysis carried out on various few-layer graphene samples confirms
that such a behavior can be ascribed to the electron-electron
correlation, which reduces the EPC strength for optical phonons near
the $K$ point and the probability of the associated Raman process.
Therefore, we have reported a clear signature of the
electron-electron interactions in multilayer graphene, and shown an
experimental evidence of the EPC tuning which can be easily achieved
by a simple chemical doping method.

\end{document}